# The Internet of People: A human and data-centric paradigm for the Next Generation Internet

**Marco Conti, Andrea Passarella**

Italian National Research Council, IIT Institute, Pisa, Italy

**Abstract**: The cyber-physical convergence, the fast expansion of the Internet at its edge, and tighter interactions between human users and their personal mobile devices push towards a data-centric *Internet* where the *human user becomes more central than ever*. We argue that this will profoundly impact primarily on the way *data* should be handled in the Next Generation Internet. It will require a radical change of the Internet data-management paradigm, from the current *platform*-centric to a *human*-centric model. In this paper we present a new paradigm for Internet data management that we name *Internet of People (IoP)* because it embeds human behaviour models in its algorithms. To this end, IoP algorithms exploit quantitative models of the humans' individual and social behaviour, from sociology, anthropology, psychology, economics, physics. IoP is not a replacement of the current Internet networking infrastructure, but it exploits legacy Internet services as (reliable) primitives to achieve end-to-end connectivity on a global-scale. In this opinion paper, we first discuss the key features of the IoP paradigm along with the underlying research issues and challenges. Then, we present emerging data-management paradigms that are anticipating IoP.

## 1. Introduction

The Internet is expanding at an exponential pace, thanks to the diffusion of personal mobile and IoT devices, and the unprecedented diffusion of the pervasive Internet. Three main expansion directions can be noticed. From the standpoint of devices, this expansion *is happening mostly at the edges of the Internet*, rather than at its *core* infrastructure. Users' personal and IoT devices already outnumber core devices, and this trend is not going to stop anytime soon ([N16][C17]). A complementary trend coupled with the expansion of the Internet at the edge is the migration of network and computing functionalities towards the edge ([SPC09], [BMZ12], [FLR2013], [KOM14], [HAHZ15], [LMED15], [YLL15], [BBCM16], [CZ16], [MC2016], [AD17], [MMG17], [RMMS17], [VPC17]).

The second dimension is related to data. The Internet, particularly at its edge, is becoming primarily (even though not exclusively) a *data-centric network* ([KCC07], [TP12], [SKS14], [XVS14], [ZABJ2014], [ABCM16]), in the sense that – more and more – users exploit the Internet to access data, more than to connect to specific devices, and their devices produce and are constantly exposed to huge amounts of data generated by other connected devices. Thirdly, the Internet is expanding into the physical world, or, better, the boundaries between the physical world and the cyber world (of Internet and Internet applications) are more and more blurred. This generates a *Cyber-Physical Convergence* where data flow between the physical and cyber dimensions, and actions in one of them impact immediately on the other ([CDB12], [XHL14] [SV15], [CPAMM2016], [SKEHBC2016]). Examples can be observed in the control of critical infrastructures through networks of IoT devices, or in "real" human activities in the physical world, e.g., shopping, voting, working, traveling, etc., ([IKM10], [RLSS10], [BR2012], [JGMP2014], [LBK2015], [AFSAK2016], [C16], [ZDWZ2016], [BAR17], [CE17], [MB17], [QMA17]) highly influenced by information gathered from "cyber" services (e.g., from online social networks [FPQSS16], [BM17], [CP17]).

Along all these dimensions, *humans are bound to play a key role in the Next Generation Internet paradigms*, much more important than it has ever been in the current (and past) Internet ([CCFJK2011], [CBKM2015]). Humans have a direct relationship with devices at the edge, because either they are their own personal devices, or they instrument the physical environment where users live. Humans, through their personal devices, are constantly exposed to (and contribute to generate) the huge amount of data that reside in the Internet. Humans impact on (and are significantly impacted by) the actions of the different actors of the cyber-physical world. Humans and the Internet devices, through which they communicate, become actors of a complex socio-technical ecosystem. One of the most intriguing effects of this convergence is that the *human becomes the center of the Internet system* and, for this reason, in [R2009], this paradigm change has been termed an "Anti-Copernican Revolution".

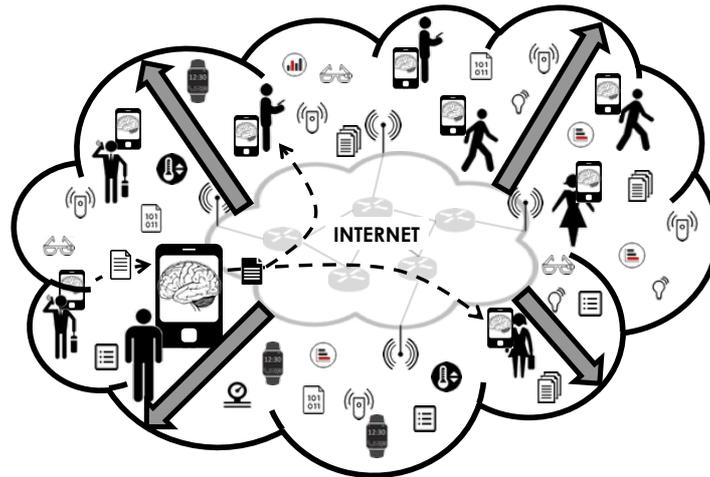

**Figure 1. Cyber-physical convergent scenario for the Next Generation Internet**: edge devices prominence, human centricity, data centricity.

Figure 1 depicts this scenario, where human users have plenty of connectivity opportunities either through the core Internet, with other users in proximity through self-organizing networking, or with physical devices through IoT technologies (or combinations thereof) ([AIM2010], [AIMN12], [MSDC2012], [BCGS2013], [AGM15], [CBKM2015], [BGLLP2016]). Even more importantly than connectivity, users are immersed in a huge amount of data that they could in principle access at any point in time. Data may come from remote users and (IoT) devices, but most of the time data will have a strong locality dimension, and will enable interactions between users nearby, and between users and "local" things ([GZW13], [ACD14], [PZC14], [DAC2016]).

Among the many issues posed by these emerging trends, the *way data will be managed* in a converging cyber-physical world will be of paramount importance for the way users access them and all the services based on them. The increasingly pronounced data-centric character of the Internet is in fact making data management at least as important as the underlying networking technologies. However, in the *current-Internet* data-management paradigm, users have to rely on global platforms (e.g., Facebook, Google, blogs, news sites, etc.) to explore, filter, and obtain data of interest. In the perspective of the Next Generation Internet described so far, this approach presents key roadblocks that are already emerging more and more clearly, such as

- lack of trust in the obtained information;
- lack of transparent privacy policies, configurable and under the individual user's control;
- constant monitoring of users' behaviour by global platforms to provide to them "navigation" and filtering services to find relevant data embedded in the huge amount of available data;

- loosing precious relevant and trusted information in the middle of huge quantities of irrelevant "noise" (*false negative*), or being exposed too frequently to irrelevant information (*false positive*);
- technical inefficiencies: although, more and more, information relevant to a user is local, i.e., available on other devices nearby, it needs to be brought back and forth between users physically nearby and global centralized platforms.

We argue that these issues arise, by and large, by the exclusive use of platform-centric, rather than human-centric, paradigms according to which Internet data management (and Internet more broadly) has been conceived so far. To address the above issues, a *radically new human-centric approach to Internet data and knowledge management*, which we call the Internet of People (IoP), is emerging.

IoP derives from research ideas recently described in ([Q16], [CPD17], [L18],) and around which an international research community is gathering [B17]. Considering the scenario depicted in Figure 1, a first fundamental characteristic of IoP is that it *radically departs from the current Internet data-management paradigms*, based almost exclusively on huge, but "distant from the user" global platforms. IoP will exploit such platforms when needed, but it will turn the data management principles upside-down, *placing the human* (and their personal devices) *at the centre of the data-management design* [CPD17]. Along a similar line, IoP will not be a replacement for the current (or next generation) Internet. IoP data-management functions will work *on top* of the legacy-Internet networking services, dynamically selecting the most suitable one for communication, without changing them.

A second key characteristic of IoP is that *users' personal mobile devices assume a very special role*. As they are the "gateways" through which users access the converging cyber-physical world, they become the *proxies of their human users in the cyber world*. Therefore, in IoP, users' personal devices are not anymore passive generators and consumers of data, but they *play an active role in data management*, either through local decisions, or through collaborative decisions with other devices with which they interact. IoP will thus be (humans') device centric, in the sense that users' devices will be primarily responsible to autonomously build and configure the data management services they require, instead of delegating these tasks to remote centralized platforms.

A third fundamental characteristic of IoP will be that, in doing so, users' devices will *incorporate models of their human users' behaviour*. Humans, through the evolution process, have developed effective methods to select relevant information among huge amounts of data. In IoP we wish to *embed these capabilities* in the data-management algorithms running on personal devices, as they are the proxies of the humans in the cyber world. To this end, IoP devices will use *quantitative models* describing their users' individual and social behaviour, in the form of mathematical models or algorithmic descriptions. They will be used *as-they-are* in IoP devices to replicate the *very same behaviours of their own users*. Thanks to this approach, IoP will be inherently a *human-centric* Internet paradigm.

Note that this fundamentally differentiates IoP from conventional bio-inspired paradigms [DA10]. In IoP, human behavioural models are embedded into algorithms running on users' devices, as the latter need to behave exactly as they human users would if they had to take the same data management choices. On the other hand, most conventional bio-inspired approaches design networking or data-management algorithms by exploiting analogies with natural systems. However, there is not necessarily a strong and clear link between devices in the "cyber" world, and corresponding entities in the natural system.

Figure 2 provides a few concrete examples of foreseen IoP functions. In case (a), while a user moves in the physical world, their device fetches from the cyber world only data relevant for the user at that moment in time (most often, available on nearby devices). To assess relevance, the device might exploit cognitive heuristics (i.e., models of human's brain functions derived in cognitive psychology) for relevance assessment [G11].[1] In case (b), a users' device estimates the diffusion that a piece of locally generated information would have among the members of its user's social network taking into consideration the strength of social links. To achieve this, the user's device would exploit models of human social relationships described, for instance, in [APCD15]. In case (c), in order to extract knowledge out of the data available on other devices, the user's device exploits cognitive heuristics to efficiently prune part of the available data and then uses compact representations of Deep Neural Networks to analyse in detail the remaining data [R18].

---

[1] An example of the use of cognitive heuristics for data dissemination in self-organising networks is presented in [CMP13].

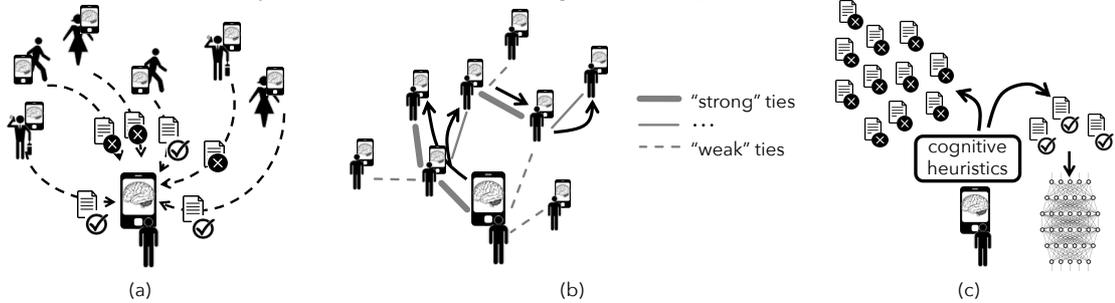

Figure 2. Examples of IoP human-centric functions.

We foresee that the IoP paradigm could be realised through a data-centric overlay on top of the Internet networking services. While data-centric overlays have been proposed many times, the primitives and algorithms to realise their functions is what distinguishes IoP. Specifically, as described in detail in Section 2 (see Figure 4), IoP would define a set of primitives, and a corresponding set of algorithms, through which applications will access the novel data-management functionalities. The definition of the primitives and the algorithms will incorporate the human-, data-, and device-centric concepts described so far. Therefore, we think that most of the IoP research challenges will be in the identification of appropriate abstractions, models and mechanisms to embed these dimensions in the IoP primitives and algorithms.

In the rest of the paper, we first highlight the key research challenges ahead for IoP (Section 2). We then highlight existing results that, in retrospect, can be seen as early instances of IoP algorithms (Section 3). Finally, in Section 4 we summarise the main characteristics of IoP and provide a discussion on the way ahead towards its realisation.

## 2. IoP Research Challenges

While some studies exist in the literature that anticipate IoP concepts (they will be discussed in Section 3) several research challenges have still to be addressed to arrive at completely defining the IoP paradigm. We start discussing the definition of an IoP framework and the human-centric design of its data-management primitives. Then, we focus on specific IoP data-management aspects such as data collection and analytics services, policies and systems for privacy preserving data handling, as well as the overall management of resources in the IoP ecosystem. We also discuss suitable models of human behaviour to be embedded in IoP data-management algorithms.

### 2.1 IoP framework

IoP operates like an overlay on top of legacy Internet networking services, thus exploiting the legacy Internet communication services that connect any node on a global scale. The IoP overlay is characterized by two main abstractions: the *IoP graph* and the *IoP primitives*.

### 2.1.1 IoP graph

The IoP graph describes the characteristics of the IoP devices, and their mutual relationships. The nodes of the IoP graph represent user devices, physical objects ("things") and, in a long-term view, also human users.[2] An edge in the IoP graph models a "data channel" between two nodes (i.e., a channel through which the two nodes can exchange data), and the properties of the "data channel". For example, in case of nodes representing user devices, the channel properties include the properties of the social relationships between the users, which can be used to estimate, e.g., the trustworthiness or priority of exchanged data. More precisely, the IoP graph can be seen as a *multilayer* graph, where each layer models a specific type of connectivity among the nodes. Specifically, as shown in Figure 3.a, Layer 0 represents the legacy Internet connectivity, i.e., the Internet routing graph where nodes are connected to the backbone through 1-hop wired and wireless infrastructure-based access networks. In this case a multi-hop path between a pair of nodes represents the possibility to exchange data between the two nodes by exploiting any suitable legacy Internet primitive. As shown in Figure 3.a, some physical objects (things) spread in the environment have no connection to Internet and hence they are not able to exchange data with other nodes by exploiting the Layer 0 of the IoP graph.

---

[2] For example, see the social computer paradigm [RG13].

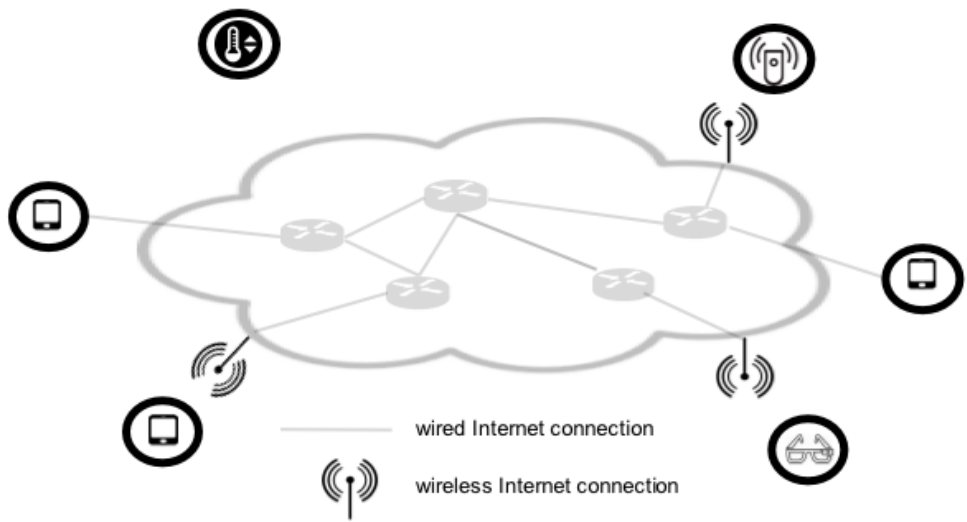

Figure 3.a: Layer 0 of the IoP graph: the Internet connectivity

The next layer, Layer 1, is obtained by adding device-to-device wireless links. Specifically, a device can communicate directly with the other devices at a one-hop distance through their wireless interfaces. By exploiting the device-to-device links, a device can be connected to the Internet by exploiting the multi-hop ad hoc network paradigm. As shown in Figure 3.b, by exploiting device-to-device wireless links, the things, which by considering only the Layer 0 graph were not connected to the Internet backbone, have now a 2-hop path to the Internet backbone and hence can communicate with all other nodes. Note that this aspect of Layer 1 is already considered in the 5G vision since the beginning [5G16]. Specifically, technologies such as LTE-D2D are expected to be massively used to guarantee range extension with respect to pure infrastructure-based cellular technologies.

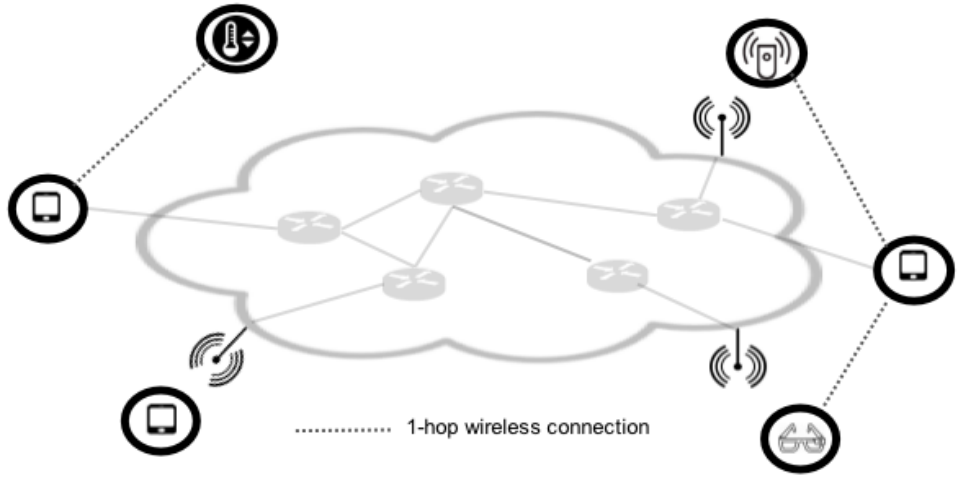

Figure 3.b: Layer 1 of the IoP graph: device-to-device connectivity

The next layer, Layer 2, is obtained by establishing the associations between devices and humans, see Figure 3.c. The associations can be either permanent or temporary. Permanent associations represent (for example) the association between a human and its personal devices, while a temporary association is the association between a human and a thing/device (e.g., a sensor, a printer, etc.) available in the environment that they are temporarily using.

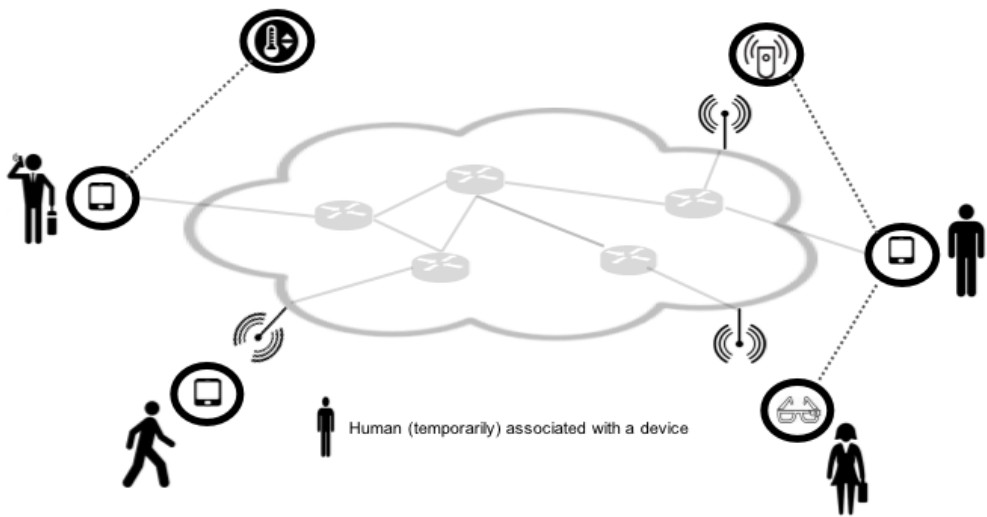

Figure 3.c: Layer 2 of the IoP graph: human-device association

A further layer, Layer 3, represents the "social connectivity" layer, e.g., a link exists between two nodes of the IoP graph if they have a social relationship. More precisely, as shown in Figure 3.d, the nodes of the IoP graph inherit the social relationships which exist among the humans that are (temporarily) associated with those nodes. For example, the social link which exists between the man with the bag and the walking man is inherited by their personal devices (i.e., the thin red link). Similarly, the weak social link which exists between the walking man and the lady is inherited by the devices they are (temporarily) associated with. Specifically, the man has a permanent association with its personal device, while the lady has a temporary association with a thing available in the environment.

Note that, thanks to this abstraction, relationships between objects in the IoP graph can "borrow" properties of the relationships between the owners of the interacting objects. From this standpoint, IoP incorporates and extends concepts such as the Social IoT ([AIMN12], **[**BYZ12], [BZWYZ13]) and Social Networks of Sensors [TMZ17].

At Layer 3 edge represents a logical channel among the nodes it connects. The data exchange between the two nodes is achieved by composing the legacy Internet primitive with IoP primitives that exploit the properties of the social relationships between the nodes (e.g., an IoP data filtering primitive based on the trust level among the nodes [AH00], [JIB07] [DC18]).

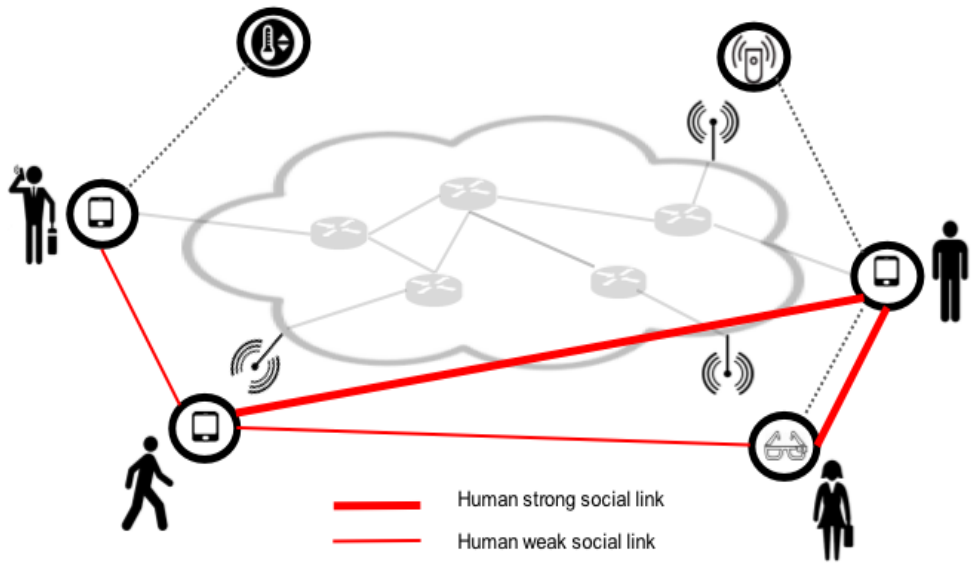

Figure 3.d: Layer 3 of the IoP graph: the social connectivity

The IoP graph is dynamic, to represent physical movements of nodes (and, therefore, opportunities of exchanging data with different sets of nodes at different times), changing relationships between users (new social relationships being established, or old one being faded away), or changing relationships with physical objects. For example, as shown in Figure 3.e, the lady, while moving in the environment, changes her associations with the physical objects and, consequently, her social relationships are inherited by the object she is currently associated with.

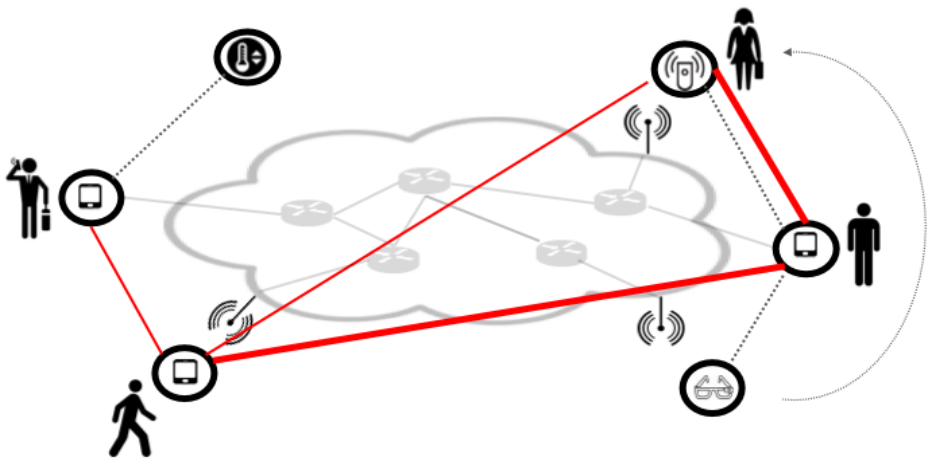

Figure 3.e: Changes in the devices social connectivity due to changes in the human-thing association

**As a consequence of this** multi-layer **representation, data** would **diffuse over the IoP graph as a function of the nodes' and edges' properties at the various layers of the graph.** However, the IoP nodes might have a more or less complete knowledge about the IoP graph. For example, it is likely that IoP nodes will have only a partial/imprecise knowledge of the entire IoP graph, either because other nodes will disclose to them only partial information, or simply because nodes might not need to know the details of the entire IoP graph.

An important aspect related to the IoP graph is how to characterise and model its properties. The goal would not only be modelling the IoP graph per se, and characterising the behaviour of IoP algorithms as a function of its properties. Ideally, these models should also provide tools to be embedded in the IoP protocols operations, indicating, e.g., optimal configuration of the protocols based on the IoP graph properties, or ways to modify the IoP graph to achieve a target behaviour in terms of data management. The characterisation of the IoP graph is itself a "multilayer" problem. On the one hand, it would be important to have ways to characterise the properties of the IoP graph at a global scale, considering complete knowledge of the graph. This would be important to assess global IoP properties, such as the overall robustness of the graph, the presence of particularly important or critical nodes, the overall properties of information diffusion. On the other hand, it would also be important to characterise local properties of the graph, which could be the main properties determining local diffusion of information or sharing of resources between nearby nodes. To this end, we expect that physics of complex systems will play a significant role. Models already exist in this literature that address exactly these challenges, for different types of graphs ([C07], [DM03]).

The elegance, and powerfulness, of these models is that they typically provide compact mathematical models that capture in simple forms very complex global properties of the graph. Moreover, significant developments are emerging in this literature about graph reconstruction techniques, or techniques to characterise graph properties under partial or uncertain knowledge, which would be clearly very useful in IoP [C12, C15]. Another significant area from this standpoint are control-theoretic models, which describe the properties of complex socio-technical phenomenon over graphs through control-theoretic formalisms [PRD18]. The importance of this area is that they not only explain the emergence of certain phenomena over complex graphs, such as the diffusion of opinions in a social network. They also provide practical tools to control or influence these processes, by tuning the properties of the graph.

### 2.1.2 IoP primitives

The *IoP primitives* will be the *data-centric* primitives that define the type of interactions between the nodes of the IoP graph. We envisage two "layers" of primitives, as illustrated in Figure 4. The first layer provides the IoP "basic" primitives, while the second layer provides IoP "enhanced" primitives, obtained by appropriately combining the basic primitives. The main rationale for this separation is to identify, as the basic primitives, a minimal set of primitives that could support all required IoP data-management functions, and allow for evolution of IoP by the definition of novel, today unforeseen, functionalities. Such functionalities might be encapsulated in enhanced primitives or be implemented directly in the applications. Applications could thus access both basic and enhanced primitives. On the other hand, both set of primitives would exploit legacy-Internet communication primitives to move data between nodes.

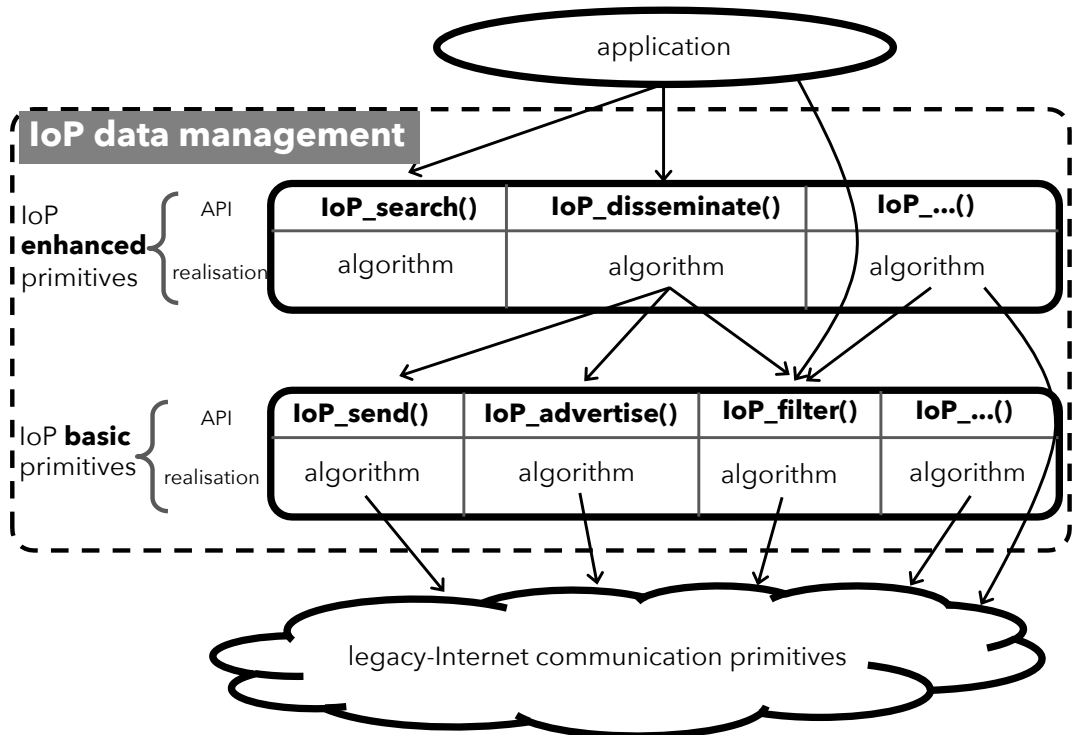

**Figure 4. IoP primitives (architectural view).**

Note that this separation between basic and extended primitives is very similar to the way Internet has been conceived and evolved over time. Also in the legacy Internet we can identify, in retrospect, a set of "basic" primitives (such as unicast, multicast, broadcast, connection-less and connection-oriented communications), on which richer and richer functionalities have been defined over time, through appropriate combinations of them. Initially, Internet applications have been designed directly "on top" of the basic primitives. Over time, functionalities have been encapsulated in "enhanced" primitives (or services), provided as additional functions to applications.

In any case, IoP primitives will be human-centric by exploiting the models of how humans access and filter data, as well as how they share, advertise and trust information exchanged between each other. Moreover, IoP primitives must embed trustworthiness in data access and in personal interactions, as well as implement privacy preserving policies for the personal data.

It is hard to tell upfront which functions should be included in the basic primitives and which in the enhanced primitives. Drawing such a line between the two sets of primitives is one of the challenges ahead for IoP research.

Let us provide an example of what a basic and what an enhanced primitive could be. Figure 5 presents the case of a possible basic IoP primitive that a user can invoke for sending data with a controllable level of trust to two other users of the IoP graph, where the level of *Social Closeness* (SC) [APCD15] among human users can be used as a proxy for the level of trust among them.

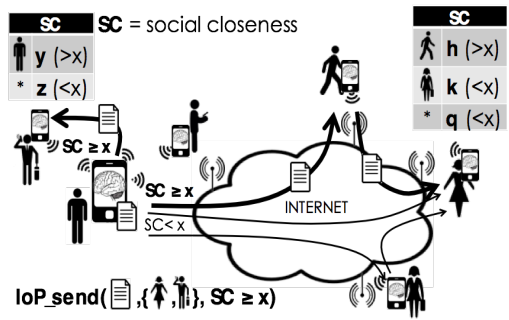

**Figure 5. Trusted data exchange IoP primitive.**

Specifically, the primitive "**IoP_send**(*data*, {set of receivers}, SC ≥ *x*)" identifies the paths over the IoP graph, between the sender and the designated receivers, such that *data* arrives at each receiver by crossing only edges with a level of SC ≥ *x*. This prevents that a node in the path discards *data* because it has received it from a node that it does not trust sufficiently (i.e., with level of SC < *x*). The level of SC is a property of the social link (in the sense of Figures 3d and 3e), which models the trust between the IoP nodes connected through that link. Nodes estimate the SC of their links (more precisely, the social link between their users) by exploiting reference models of the structure of human social relationships, such as the ego-network models [ZSHD05, HD03].

As said before, the IoP graph is a multilayer graph, where different types of links may exist among the users. In Figure 5 one destination is reached directly through a 1-hop ad hoc network, as the SC level among the nodes connected through that link is enough to satisfy the constraint on the trust level. On the other hand, the other receiver has to be reached through an intermediate IoP node. This node "validates" the data before forwarding it to the final destination. In this example, the IoP nodes communicate over conventional multi-hop Internet paths, with state-of-the-art encryption, if needed. Thus, the example also illustrates how legacy Internet primitives can be used in IoP.

Data filtering could be another basic primitive that IoP should provide. Other IoP basic primitives could deal with supporting access to data, irrespective of the location where it is generated/stored, exchange data and information with other users based on their specific needs and actions (which may change dynamically over time), or to guarantee data privacy.

On the other hand, data dissemination could be an example of an enhanced IoP primitive. In this case, data dissemination would combine at least three basic primitives. An "advertisement" primitive, through which nodes can become aware of which other nodes are interested in certain data. The "IoP send" primitive, illustrated in Figure 5, to send data from the originator node to the interested nodes. And a "filtering" primitive through which receiver nodes can further filter the received data. Other similar examples of enhanced primitives could include replication of data to support its availability and access to it (possibly within specified time bounds), search for data, etc.

**2.2 IoP data management and analytics**
The *IoP data-management algorithms* are a set of protocols, used by nodes in the IoP graph, to implement the semantic of the IoP primitives (both basic and enhanced), as today's Internet nodes implement the routing and forwarding protocols to support the unicast, multicast, anycast communication primitives, as well as higher layer services such as content-centric communications and caching.

We envisage algorithms of (at least) four different classes, i.e., (i) "reactive" algorithms, through which devices will filter data they will be exposed to, without explicitly asking for it, e.g., see Example (a) in Figure 2 (ii) "proactive" algorithms, through which nodes will advertise availability of data or explicitly look for them, (iii) data dissemination algorithms to spread information to all interested nodes in the IoP graph, e.g., see Example (b) in Figure 2 and (iv) "data analytics" algorithms, through which nodes will process data to extract relevant knowledge, e.g., see Example (c) in Figure 2.

While the definition of specific algorithms is clearly out of the scope of this paper, it is possible to identify some key design guidelines for the algorithms, to make them compatible with the IoP concepts described in Section 1. IoP data-management and analytics algorithms will be built on a heterogeneous set of resources, available in the IoP graph, provided by devices owned by specific users, devices not bound to any specific user, and human users themselves. In addition, IoP algorithms should be designed such that users' devices are in control of the behaviour of the data management algorithms, and that they implement the algorithms as "proxies" of their human users. Hence, key questions to be addressed include:
  i) Which are the human behavioural models to be exploited in the design of these algorithms?
  ii) What is the correct approach to represent the resources provided by nodes, advertise and orchestrate them, and make them available in the IoP ecosystem to build complex network functions?
  iii) How to realize privacy preserving algorithms?
  iv) How to design effective and efficient algorithms taking into account that nodes might have only partial or imprecise knowledge of the IoP graph?

In the next subsection we provide some research directions to address the above questions.

**2.2.1 Human-centric models**
As said before, human behavioural models have to be exploited in the design of IoP algorithms. Adopting such an inter-disciplinary approach is not an easy task, as it requires to bridge very different scientific disciplines, such as: social sciences, cognitive psychology, complex networks, and microeconomics.

*Social sciences* model the way users establish social relationships, how they trust each other, and how they are prepared to share resources with each other. *Cognitive psychology* describes, among others, how human beings perceive and interact with data, how they assess relevance of information, how they exchange it when interacting, and how they extract knowledge out of it. Data-centric Internet systems for mobile networks have already been proposed (see also Section 3.1), where these models are exploited to efficiently guide information diffusion among users [MVCP16]. Very useful models have been derived in the area of *complex network analysis* [C07], describing, for example, human social relationships with compact graph descriptions, amenable to characterize human behavioural properties and exploit them in the design of networking solutions. Finally, *microeconomics* models describe how humans negotiate the use of infrastructure and content resources, trade and share them [FF02, FG07]. While these models will be developed and validated with an interdisciplinary work led by non-ICT communities, ICT technologies, like Big Data Analytics and Machine Learning, will have a major role in validating and tuning these models in the IoP context [APCD15].

**2.2.2 Resources management and trust models**
Another challenging research question is what the correct approach is to represent the resources provided by IoP nodes, advertise and orchestrate them, and make them available in the IoP ecosystem to build complex network functions. In IoP, resources may be very dynamic and heterogeneous, and only partly controlled by (more or less) trusted operators. On the other hand, it will be in the interest of all users that IoP is an ecosystem where own resources can be shared in a fair way, in return of better data-centric services built thanks to these shared pools of resources. In this perspective, it is an open challenge to identify which models should IoP adopt in order to establish trust and facilitate cooperation between parties, so that users will be confident in sharing and using each other's resources ([AH00] [JIB07] [GCT17]). To model resource availability and sharing, and the trust associated with it, we can use models derived from micro-economics [FF02, CF06, FG07]. Interestingly, differently from conventional game-theoretical models, these approaches take into consideration the effect of social relationships in the way cooperation among humans develops. It has been shown that patterns of cooperation between individuals, which determine sharing of resources, are not correctly modelled by considering only the actors' self-interest (as in conventional game theoretical approaches) but need to be modelled taking also the social dimension into account. This literature is important for IoP, as these models will be used to describe the possibility to access shared resources from the standpoint of individual nodes. For example, if a resource is contributed by other users nearby or by devices of users with whom an ego has some type of social relationship, the possibility of exploiting that resource in IoP would be described through these models. Similarly, human-centric relationships could also be exploited to design effective incentive schemes for collaboration between users through their personal devices.

**2.2.3 Data privacy**
In addition to collecting information from the cyber world around us, our personal devices, moving along with us, leave in the virtual world digital traces of our behaviour such as: our mobility patterns (e.g., our movements in the city), our social relations (e.g., the people we meet), our opinions, our resource consumption patterns, our economic/financial behaviours, and so on (e.g., [CMJ11], [GNP11], [ZZC12], [VNNR2015], [CFT2016], [OVSBA2016], [CCT17]). The big data that encode our digital traces have a huge value from the commercial, social and scientific standpoints ([ZCW14], [PPSG15], [FSSST2016], [MCG2016], [PTLG16], [SZOT2016], [CL17], [EPRF17], [RWP17]). They can be used by decision makers to provide better services to citizens but, at the same time, if not managed in an appropriate way, may compromise our privacy ([LZD11], [HDMP15], [RRFC15], [FCJ17]). Therefore, data privacy is a fundamental requirement in designing the Internet of the future ([RGK11], [BHAJ2016], [CRB2016], [RAL17], [ZLLJ17]), and in particular in the IoP design.

Therefore, fundamental components of the IoP algorithms will be techniques based on novel paradigms to *guarantee the privacy of exchanged data*, and *guarantee compliance* of data management with reference technical (such as privacy-by-design) and legal (such as the new European GDPR) frameworks. Fundamentally, also the IoP privacy models must be user centric, as users must be able to define and control the level of access to their data by third party services. This can be achieved either by storing the personal data in the cloud and controlling the access to this information [DFL16], [SB17], also using cryptographic mechanisms ([CDF10], [CS10] [KL17], [SLS17], [SYL17]), or by exploiting emerging approaches such as Personal Data Storage (PDS) [M14] and Databox [HHC15] systems. A PDS (or a Databox) is a user-controlled SW entity (typically implemented in physical devices owned by the user), which is responsible for intermediating all accesses to users' data, and elaborations of it. Applications have to interact with the PDS to get access to data, and are allowed, if the case, to install modules for data elaboration. Data always remain under the control of the user through the PDS. Any export of personal data must hence be authorized by the individual, who in turn can be compensated for the use of the data.

**3. First Steps towards IoP**

While the current Internet is natively neither data-, nor human-, nor device-centric, in the scientific literature we are observing the emergence of networking and data management paradigms that can be considered as precursors of IoP. At the network level, people-centric networking and sensing systems represent a first step in the design of data-, human-, and device-centric paradigms. In these paradigms the human personal devices have an active role and the human behaviour strongly affects the design of effective protocols which implement these paradigms. Moreover, the protocols implemented in these systems are data centric.

In people-centric networking our personal devices, typically the smartphones, via their wireless interfaces, such as LTE, 5G, WiFi, Bluetooth, or LR-WPANs ([BCG02], [MC13], [ARS16], [BBBK2016], [DBA16]), connect us directly to the devices of the other people that are nearby, or to the physical objects located in the physical space around us. Typically, we use these one-hop direct connections to share data with nearby devices that have a special value for the space-time context in which such data is exchanged (e.g., [ACD14], [YL16], [ACD17], [CD17], [DHK17], [G18] [MJM17]). Our smartphones allow us not only to collect information around us but also carry and forward information to other people we meet through various forms of self-organizing networking [CG2013]. In self-organizing networks, the movement of people, and hence of their personal devices, "create" the multi-hop paths through which the information circulates in the cyber world.

The people-centric sensing (also known as *crowdsensing* or *participatory sensing*) paradigm combines wireless communications and sensor networks with human's daily life activities to sense the physical world by exploiting the billions of connected users' mobile devices/phones, without deploying an additional sensor network ([C08], [FRNDD12], [SAS12], [HCCL13], [MZY14], [GWYW15], [RDP16], [CGF17]). A mobile phone, though not built specifically for sensing, can in fact readily function as a sophisticated sensor by exploiting the camera (as video and image sensors), the microphone (as an acoustic sensor), the embedded GPS receivers (to sense location information), etc. Other embedded sensors such as gyroscopes, accelerometers and proximity sensors can collectively be used to estimate useful contextual information (e.g., whether the user is walking or traveling on a bicycle). Therefore, humans with their smart devices during their daily activities represent potential sensing devices distributed across the physical space, which can be "enrolled" in the sensing task via opportunistic or participatory techniques ([CELM06], [LEMM08], [E10], [MZY14], [LDPX16], [SCH16], [TSHDW17]).

People-centric sensing is a human-centric paradigm for sensing the physical world. Sensed data is used in the cyber wolrds to build a virtual representation of the physical word that can be used to provide better services to the citizens. Smartphones are the reference devices for people-centric sensing, however, this paradigm can be extended to include other forms of people-centric sensing, such as vehicular sensing. Vehicular sensing exploits the large set of sensors embedded in vehicles, the personal smartphones of the drivers and passengers, and also ad-hoc sensors (e.g., for environmental monitoring (**[**HFZC16**]**) installed on vehicles to collect data at urban scale ([A07], [EGHN08], [SBS17], [YDL17]).

A special case of people-centric sensing is based on exploiting the posts done by people on online social networks as a sort of virtual sensing measures. This can complement sensing in the physical world [SOM13], [STP17], [LMP18], provided that we are able to predict the locations of the posts [BSM10], [HGS12], [HCB14]**,** [WGD17]. For example, if people in the same physical location, independently and at the same time, write posts complaining about high temperatures, this can be an indication of a particularly high temperature in the given location.

People-centric sensing is an example of a service that can be implemented by exploiting the personal-device resources. In ([CK10], [MYCH10], [PKOC12], [UKPC14], [MCP18]) this paradigm is extended by defining the opportunistic computing paradigm to exploit and coordinate all nearby personal-device resources, as well as other resources available in the physical environment. For example, a device may offer as a service its high-speed Internet connection to other devices nearby. Similarly, the possibility to offload a computational task from a resource-limited device to a nearby resource-rich device is also a research direction falling in this domain ([FGP2016] [MZAD16], [SAZN12]).

Similar to opportunistic computing, other paradigms, such as fog computing ([BMZ12] [YLL15]) and cloudlet architectures [SBCD09], [SPC09] are pushing the intelligence towards the edge of the network by exploiting gateways, at the boundary between the access network and the Internet, with a goal to provide more reactive and personalized services to the users. However, opportunistic computing represents the first paradigm that is pushing directly the intelligence up to the users' personal devices. Specifically, mobile smart devices, by pooling their resources, can start offering services as a sort of *mobile clouds* bringing services and resources closer to where they are needed [LMED15], thereby avoiding the bandwidth and management costs associated with accessing the services in the cloud [BBCM16], [CMP15].

As the focus of this paper is on data management, hereafter we do not further discuss the people-centric networking, sensing and computing paradigms and we refer the interested reader to [CBKM2015], [CPD17]. On the other hand, hereafter we present and discuss two paradigms for data collection and data dissemination which are data-, human-, and device-centric which provide interesting ideas for developing data management algorithms in IoP.

**3.1 Human-centric data collection**

In the cyber-physical converged world, the active user participation in the process of data creation and diffusion creates a huge quantity of pervasive information stored in the personal devices around us. As discussed before, in such a context, personal devices could act as the *avatars* (or proxies) of their respective users. They allow their owners to explore congested cyber information landscapes by collecting the available information, filtering it, and presenting relevant data to the human brain of their owners. It is, in the end, the brain of the user the final recipient of the collected information. At personal devices, a key challenge is therefore how to select, in an effective way, the information to present to the users, avoiding to flood them with huge amounts of (useless) information that the human brain cannot handle, with the risk that useful information gets unnoticed in the middle of a flow of irrelevant data, and discarded. In [CMP13] a methodology is proposed to overcome this problem by directly embedding in the personal devices the cognitive processes used by the human brain to filter out irrelevant information. In other worlds, our personal devices must learn how to operate in the virtual world for selecting what is important for us, as our brain would do. Specifically, the human brain performs this task using so-called cognitive heuristics, i.e., simple, rapid, yet very effective schemes to swiftly assess the relevance of information under partial knowledge ([GG96], [GG02]). Cognitive heuristics are fast, frugal and adaptive strategies of the brain that allow humans to face complex situations by addressing simpler problems using only partial information related to them [MGG10]. By exploiting cognitive heuristics, the human brain is able to swiftly contextualize the stimuli it is subject to, identify the relevant features and knowledge to be considered, assert the relevance of received data, and finally select the most useful data, even when only partial information is available. Hence, despite their simplicity, cognitive heuristics are indispensable psychological tools that are very effective in solving decision-making problems like information selection and acquisition [MGG10].

The use of cognitive heuristics to develop effective algorithms for data collection in the cyber-physical world has been first proposed in [CMP13], where two of the several cognitive models present in the cognitive psychology literature are considered ([GG96] [G08] [MGSG10]): the *Recognition Heuristic* and the *Take the Best Heuristic.*

The *Recognition* heuristic assumes that merely *recognizing* an object is sufficient to determine its relevance [GG02], where an object is recognized if the user "sees" it a sufficient number of times[3]. Therefore, in [CMP13] the recognition-heuristic principle is exploited to let each personal device rapidly decide the utility of taking one data item available in the cyberspace, instead of another. More precisely, data are stored in the personal devices of other users moving in the same physical space and a new data is available as soon as devices are in direct contact through their wireless interfaces. It is assumed that nodes have a limited-size cache where data available at other nodes are stored. Cognitive heuristics are used to decide which subset of the encountered data should be stored in the cache. Specifically, building upon the recognition heuristic, an algorithm was proposed that is inspired by the *Take-the-Best* cognitive scheme. This algorithm uses the reference model, in the cognitive literature, of Goldstein and Gigerenzer [GG96] and exploits the recognition heuristic in order to simplify and limit the complexity of the data-selection task. The algorithm is applied whenever two nodes meet directly, and is used by a node to decide which data to store in the local cache, among the union of the data already stored locally, and the ones it could fetch from the encountered node. This is done by a recursive creation of small subsets of the available data, from which the data to be stored are extracted. The bottom-line idea is thus to rank data items based on the recognition heuristic, and store only the top ranked items (until the local cache is full). Ranking done via the recognition heuristic approximates the ranking that would be done with complete information about the available data (e.g., relevance for the local user, relevance for all the users in the local users' social communities, current diffusion of the data item). While computing the latter ranking is typically unfeasible due to resource and time constraints, ranking based on the recognition heuristic is very efficient, as it only requires local (partial) information. Note that the algorithm proposed in [CMP13] applies the very same algorithm defined in the cognitive psychology literature to describe the corresponding cognitive heuristics, thus implementing in the users' devices the very same process that their users would perform in their brain. The work presented in [CMP13] proves the suitability and effectiveness of these heuristics in problems, like data dissemination in cyber-physical world, where every node has only a partial knowledge about its environment.

---

[3] "Seeing" should be intended as a very general concept, which is not necessarily bound to visual perception. For example, the same heuristic can also be applied to concepts, e.g., which might be encountered by a person while reading a book.

Generally, human interest is characterised by assuming it is rather static and quite simple to describe [CMP13]. Users are supposed to be interested in predefined content categories (e.g., sports, movies, etc.) and therefore their devices should collect all the contents related to those categories. However, in the reality the users' interests can change over time, as a result of a knowledge acquisition process that is also the effect of social interactions between them. To model the dynamicity of users' interests based on cognitive schemes, in [CMPR13] a semantic network is introduced to represent the data stored in the users' devices. A semantic network is a reference model used in the cognitive literature to represent concepts and exchange of knowledge through interactions (e.g., discussions) between people. A semantic network is a graph, where vertices are the semantic concepts (e.g., the tags associated to data items) and edges represent the connections that exist among them (e.g., the logical link between two tags). The semantic description could be the case, for instance, of tagged photos on Flickr and Instagram, or messages annotated with hashtags in Twitter and Facebook. Note that in [CMPR13] the semantic network represents concepts, while data items stored by the personal devices are not directly represented in the semantic network. Instead, they are associated to a set of concepts, which they refer to (i.e., concepts are considered as tags, and data items can be tagged with multiple concepts). When personal devices "encounter" each other, they first exchange information extracted from the semantic network, and this drives the exchange of specific data items, as explained in the following. This process is the same behind exchange of knowledge and data associated to knowledge between human beings. Specifically, in [MVCP16] by exploiting the semantic-network representation, the selection of information to exchange starts from the concepts in common between the two nodes, similar to the way a real discussion between humans typically starts. Then, the semantic network of each personal device is navigated starting from these common concepts, and each node extracts a selection of concepts to pass over to the other node. Finally, the contents stored at the nodes, referring to the exchanged semantic concepts, are also exchanged between them. Not only this results in exchange of data, but also in an update of the nodes' semantic networks, as a side effect of the exchange of concepts between them. Therefore, users' interests (modelled through concepts in the semantic network) change dynamically, drive the dissemination of semantic data, which in turn drives the dissemination of content. The novel aspect of this work is therefore the use of cognitive heuristics to optimize the joint dissemination of semantic information and associated contents.

Cognitive heuristics have also been applied to provide cost effective algorithms for solving several other problems emerging in the cyber-physical world. Cognitive heuristics and models of dynamic memory structures have been used in [M15] for crowdsourcing in smart city environments, while in [MPC17] the social circle heuristic [PRH05] is exploited to develop a dissemination policy for opportunistic networks where each node stores data items that are relevant for itself and for other nodes in its social context.

**3.2 Human social relationships and data dissemination**
As shown in the examples presented in the part b) of Figure 2 and in Figure 5, in IoP the models of human social relationships are the basis for developing data distribution primitives with trust guarantees. Hence an important issue is to identify suitable quantitative models of the human social relationships, which can be used in the design of effective IoP primitives.

Models of the human social relationships have been extensively investigated in the social sciences literature. In [ACPD17], after reviewing one of the most promising quantitative models for characterizing human social relationships, the authors investigate how this model can be applied to design and evaluate novel data dissemination protocols in IoP. Specifically, in [ACPD17] the authors focus on the ego-network model, which has been proposed in the anthropology literature, to describe the social relationships of an individual (ego) with its social peers (alters). Ego networks are one of the key concepts to study the microscopic properties of personal social networks [APCD15]. There exist different definitions of ego networks, corresponding to different approaches in analyzing them. Hereafter we used the definition proposed in [SDBA12]: an *ego network* is formed of a single individual (*ego*) and the other users directly connected to it (*alters*).

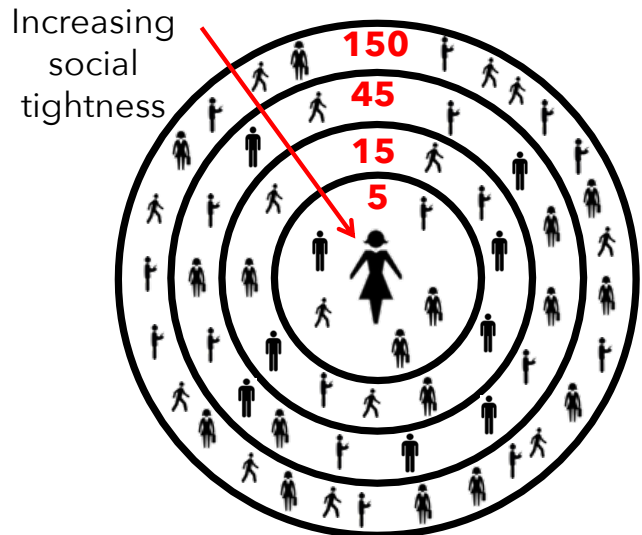

**Figure 6. Dunbar ego-network schematic representation (from [ZSHD05]). Alters are represented inside concentric layers around the ego (at the center of the picture). Membership in layers is determined by the strength of the social tie with the ego.**

According to the ego-network model, a fundamental cognitive constraint in the human's personal social network is the Dunbar's number [D93]. This is the number of relationships that an ego *actively* maintains in its network over time, i.e., the relationships for which a minimum frequency of interactions (typically, one per year) is maintained over time. The Dunbar's number in offline ego networks is known to be limited by the cognitive constraints of the human brain, and by the limited time that people can spend in socializing. In addition, it is known that cognitive constraints lead people to unevenly distribute the emotional intensity on their relationships. As depicted in Figure 6, this results in a hierarchical structure of inclusive 'social circles' of alters around the ego with characteristic size and level of tie strength (the strength of the social relationship) [LQ17]. Specifically, in the reference ego-network model [ZSHD05], there is an inner circle (called *support clique*) of 5 alters on average, which are considered the best friends of the ego. These alters are contacted at least once a week, and are the people from whom the ego seeks help in case of emotional distress or financial disaster. Then, there is a second layer of 15 alters, called *sympathy group* (which includes the support clique) containing close friends of the ego, those contacted at least once a month. After this layer, we find a group of 50 alters, called *affinity group* or *band* that contains an extended group of friends. The last circle, called *active network*, contains on average 150 alters (the Dunbar's number) contacted at least once a year. These people represent the social relationships that the ego maintains actively, spending a non-negligible amount of time and cognitive resources interacting with them so as to prevent the corresponding social relationships decaying over time. Evidence to support the existence of Dunbar's number and the described ego-network structure in offline social networks has come from a number of ethnographic and sociological sources [D93], [ZSHD05]. More recently, results have also been shown on the presence of Dunbar's number and the ego network structure in phone-call networks [MMLM13], [GQRZ17] and in online social networks [APCD15], [DAC15]. In [ALPC16], the authors have investigated the impact of the ego network structures on information diffusion in Online Social Networks by exploiting the online social links. Specifically, they have shown that, by considering the structural properties of ego networks, it is possible to accurately model information diffusion both over individual social links, as well at the entire network level, i.e., it is possible to accurately model information cascades [GHF13]. Moreover, the authors have analysed how trusted information diffuses in OSNs, assuming that the tie strength between the nodes (which, in turn, determines the structure of ego networks) is a good proxy to measure the reciprocal trust. Interestingly, they have shown that not using social links over a certain level of trust drastically limits the information spread; for example, when only very strong ties are used, the information diffuses up to only 3% of the nodes. However, inserting even a single social relationship per ego, at a level of trust below the threshold, can drastically increase information diffusion. This is consistent with the well-known Granovetter's results, that showed that strong ties can carry a significant amount of information, although weak ties are also important for acquiring diversity of information [G73]. Finally, when information diffusion is driven by trust, the average length of the shortest paths is more than twice the one obtained when all social links can be used for dissemination. Other analyses in the latter case have highlighted that in OSN also, users are separated by about 6 (or less) degrees of separation. The results presented in [ALPC16] show that when we need trustworthy "paths" to communicate in OSN, we are more than twice as far away from each other (from a purely topological standpoint).

The relevance of ego-network structure in the study of OSN properties opens up several research directions, which are relevant for IoP. The dependence of information diffusion on the trust of social relationships can have a significant impact on the design of novel social networking platforms such as Distributed Online Social Networks (DOSN), as discussed in ([G18], [KF17]). DOSN implement the functionalities of OSN platforms, but in a completely decentralized way. DOSN are human-centric and device-centric in order to maintain the control on personal data. In fact, personal data of the users and the content they exchange are stored directly on their devices, without the need of any third-party server to operate the social networking platform. This provides much more control of the user over their personal information, but requires caching and replication techniques to guarantee data availability ([CSGP14], [DDGR16], [G18]).

## 4. Discussion and Conclusions

The Internet is expanding exponentially at its edges, due to a huge number of devices through which users access the cyber world. User devices become *proxies* of their users in the cyber world: they communicate, exchange and manage data on behalf of their users, and thus should behave the way their human users would do if interacting with each other in the physical world. More and more, the physical and the cyber (Internet) worlds are blending, and what we do in one has effects on the other. The human user is thus becoming *the centre* of the Internet, which in turn is mostly becoming data-centric. This requires a radical change of Internet data management, which is now platform-centric, rather than human-centric. This brings to the emergence of a new paradigm for data management (among users' devices) in the Internet: the *Internet of People* (IoP). IoP, is a complex socio-technical system where humans have full control of their own data, and personal devices are their proxies in the cyber-world, i.e., they act as their users would do when managing data, as data are collected and elaborated for the benefit of devices' owners.

Stretching this vision further, IoP embraces even a tighter integration between the Internet and humans, allowing humans themselves to contribute (cognitive) resources to the Internet functions. In IoP humans can also be "used" as network nodes, when their role is the most suitable one for the realization of specific parts of the IoP algorithms (i.e., the way IoP primitives are realized on the IoP nodes). This is an evolution towards the social-computer vision [RG13], where the human user is perceived as another entity of the computing and communication ecosystem, whose behaviour can be modelled and predicted (clearly, up to a certain extent), and whose resources can be shared and exploited to optimize the operations of the system. As a starting rudimentary example of this vision, we may think of crowdsourcing systems, where humans are involved to solve complex problems in a synergic way together with computers ([KO14], [MZY14], [GWYW15], [KTK15]).

Therefore, we argue that in IoP we need to take into consideration the human behaviour as a *structural design paradigm*, rather than as an afterthought. To summarize, the key characteristics of the IoP paradigm are:

- *IoP is human-centric,* and, as a consequence, is *multi-disciplinary*, as IoP algorithms should be based on quantitative models of the human individual and social behaviour derived and validated in various research communities, such as sociology and anthropology ([HD03], [ZSHD05], [RD11]), cognitive psychology ([GTA99, S05, GHP11]), micro-economics [FG07, FF02, CF06], physics of complex systems (e.g., [BA99, DM03, C07]).

- *IoP is device-centric*, as users' devices are seen as "core IoP nodes", which are proxies of the humans in the cyber world, and host a significant part of the logic of the IoP algorithms.

- *IoP is data- and computing-oriented,* as IoP will naturally include primitives dealing with data management and data-centric computing, because data access, rather than connection to specific devices, is what humans will use the Internet for, most of the time.

- *IoP is also self-organizing,* as in IoP users' devices can establish spontaneous, infrastructure-less, networks with nearby devices, if local communications are the most effective ways to achieve a given task, e.g., exchanging data with the devices of other people that share the same location.

On the other hand, there are clear boundaries that define what IoP *is not*. Specifically:

- *IoP is not a replacement of the current Internet communication infrastructure.* IoP envisions that the current Internet will remain the most suitable means of global-scale end-to-end connectivity. IoP algorithms will use it as one of the enabling technologies, but will develop a radically new human-centric and device-centric data management paradigm, which will exploit global-scale networking through Internet technologies when appropriate.

- *IoP is not an application-level human-centric paradigm.* While human-centric concepts have been sometimes used to design applications, IoP is not about defining an application-level paradigm. IoP uses a human-centric perspective to define a novel *data-management* Internet paradigm, on which (human-centric) applications could naturally be developed.

- *IoP is not about Internet policy.* In IoP the human-centric focus is not related to regulatory, governance or policy aspects, but it is strictly bound to the definition of the data-management primitives and algorithms. Accordingly, IoP seeks quantitative, and not qualitative, models of the human behaviour, as only the former can be directly embedded "as-they-are" in the IoP primitives and algorithms.

In this paper we have briefly discussed the key elements of IoP: the IoP graph, the IoP primitives and the algorithms required to implement them. The personal devices, are the key nodes of the IoP graph, an overlay built on top of the legacy Internet to implement a human- and data-centric paradigm. On the other hand, IoP primitives will naturally be data-centric and deal primarily with data-management issues, relying on legacy-Internet primitives for global communication. The paper also discusses an initial conceptual architecture highlighting the relationships between Internet networking services, IoP primitives, and applications.

A key challenge in IoP is how to embed models of the human behaviour in its algorithms. To achieve this, we need to link our traditional technology-oriented perspective closely to human-centric sciences (describing various facets of the humans' behaviour) for designing IoP networking and data exchange mechanisms that are human-centric. Models of the human behaviour has to be embedded into the IoP protocols and devices logic, to drive their operations. We believe that a cornerstone to fruitfully follow this approach is to seek *quantitative* mathematical models (rather than qualitative descriptive models) or algorithmic definitions to describe the needed facets of the human behaviour.

We remark again that the proposed human-centric approach to IoP is not yet another bio-inspired networking design wave [DA10]. Due to the fact that user devices act as proxies of their users, and the human brain is often the final destination of the information collected in the Internet, embedding efficient models of the human behaviour in the core design of networking systems is a natural way to make devices behave as their human users would do if faced with the same choices and decisions.

While IoP is a radically new paradigm for data management in the Internet, some systems exist already which represent the first steps toward the realization of IoP. In particular, in the paper we have presented some existing works in the literature to design data-management protocols that embed quantitative models of the human's individual and social behaviour for designing effective data filtering and dissemination algorithms in a cyber-physical world. While some precursors of IoP exist, these systems represent only preliminary examples of application of the IoP concept, and need to be significantly refined, improved and integrated in a comprehensive IoP framework. To achieve this, several research challenges need to be addressed. This paper identified a first set of open research challenges ranging from the definition of the IoP graph and primitives, to the design of human-centric and data-centric IoP algorithms and protocols including cross layer issues, such as IoP resource management and trust models, and privacy preserving policies/systems. We have presented initial work available in the literature on some of these elements. However, until now, aspects relevant for IoP have been investigated in isolation, while IoP provides a unifying concept for their harmonic development. In addition, in the view of the new concept of IoP, each of these elements will clearly need to be reconsidered, further investigated and developed.